\documentclass[runningheads]{llncs}

\usepackage[T1, T2A]{fontenc}
\usepackage[utf8]{inputenc}
\usepackage{graphicx}
\usepackage{booktabs}
\usepackage{amsmath}
\usepackage{xcolor}
\usepackage{multirow}
\usepackage{diagbox}
\usepackage{dirtytalk}
\usepackage{academicons}
\usepackage{colortbl}
\usepackage[most]{tcolorbox}
\usepackage{enumitem}
\usepackage{xurl}
\usepackage{cite}
\usepackage{csquotes}
\usepackage{listings}
\usepackage[title]{appendix}
\usepackage{tablefootnote}
\usepackage{algorithm}
\usepackage{algpseudocode}
\usepackage{epigraph}
\usepackage{booktabs}
\usepackage{subfig}
\usepackage{float}
\usepackage[depth=3]{bookmark}
\usepackage{minitoc}
\usepackage{pifont}
\usepackage{orcidlink}

\definecolor{blue_teaming}{HTML}{0F52BA}
\definecolor{red_teaming}{HTML}{971a1e}
\definecolor{color_a}{HTML}{893244}
\definecolor{color_b}{HTML}{4F404C}
\definecolor{color_c}{HTML}{007F7F}
\definecolor{color_d}{HTML}{6082B6}

\urldef{\StyloMetrixMetricsList}\url{https://github.com/ZILiAT-NASK/StyloMetrix/blob/main/resources/metrics_list_en.md}

\title{
    Tuning for TraceTarnish: Techniques, Trends, and Testing Tangible Traits
}

\author{
    Robert Dilworth \orcidlink{0009-0005-5497-9810}
}

\authorrunning{
    Robert Dilworth
}
\titlerunning{
    Tinkering with \textsc{TraceTarnish}
}

\institute{
    Department of Computer Science and Engineering, Mississippi State University, Mississippi State, Mississippi, USA\\
    \email{rkd103@msstate.edu}
}

\hypersetup{
    pdftitle={Tuning for TraceTarnish: Techniques, Trends, and Testing Tangible Traits},
    pdfsubject={cs.CR, cs.CL, cs.IR},
    pdfauthor={Robert Dilworth},
    pdfkeywords={Adversarial Stylometry, Privacy, Unicode Steganography with Zero-Width Characters, Feature Selection Using Information Gain},
    colorlinks=true,
    linkcolor=color_a,
    citecolor=color_c,    
    urlcolor=color_d,
}

\begin{document}

\maketitle

%Note: Remove the manual hyphenation points (\-) when submitting our manuscript.
\begin{abstract}

    In this study, we more rigorously evaluated our attack script \textsc{TraceTarnish}, which leverages adversarial stylometry principles to ano\-nymize the authorship of text-based messages. To ensure the efficacy and utility of our attack, we sourced, processed, and analyzed Reddit comme\-nts---comments that were later alchemized into \textsc{TraceTarnish} data---to gain valuable insights. The transformed \textsc{TraceTarnish} data was then further augmented by \textit{StyloMetrix} to manufacture stylometric features---features that were culled using the Information Gain criterion, leaving only the most informative, predictive, and discriminative ones. Our results found that function words and function word types (L\_FUNC\_A \& L\_FUNC\_T); content words and content word types (L\_CONT\_A \& L\_CONT\_T); and the Type-Token Ratio (ST\_TYPE\_TOKEN\_RAT\-IO\_LEMMAS) yielded significant Information-Gain readings. The identified stylometric cues---function-word frequencies, content-word distributions, and the Type-Token Ratio---serve as reliable indicators of compromise (IoCs), revealing when a text has been deliberately altered to mask its true author. Similarly, these features could function as forensic beacons, alerting defenders to the presence of an adversarial stylometry attack; granted, in the absence of the original message, this signal may go largely unnoticed, as it appears to depend on a pre- and post-transformation comparison. \say{In trying to erase a trace, you often imprint a larger one.} Armed with this understanding, we framed \textsc{TraceTarnish}'s operations and outputs around these five isolated features, using them to conceptualize and implement enhancements that further strengthen the attack.

    
\end{abstract}

\section{Introduction}
\label{sec:Introduction}

    %\epigraph{\shapepar{\eyeshape}{\textcolor{red_teaming}{When I leave, burn the spread of this bed, that I touched. Burn the chair in the living room, in your wall incinerator. Wipe down the furniture with alcohol, wipe the door-knobs. Burn the throwrug in the parlour. Turn the air-conditioning on full in all the rooms and spray with moth-spray if you have it. Then, turn on your lawn sprinklers as high as they'll go and hose off the sidewalks. \textit{With any luck at all, we can kill the trail}\dots}}}{\textit{Fahrenheit 451 \\ Ray Bradbury}}
    \epigraph{    
        \includegraphics[width=1.25\linewidth]{{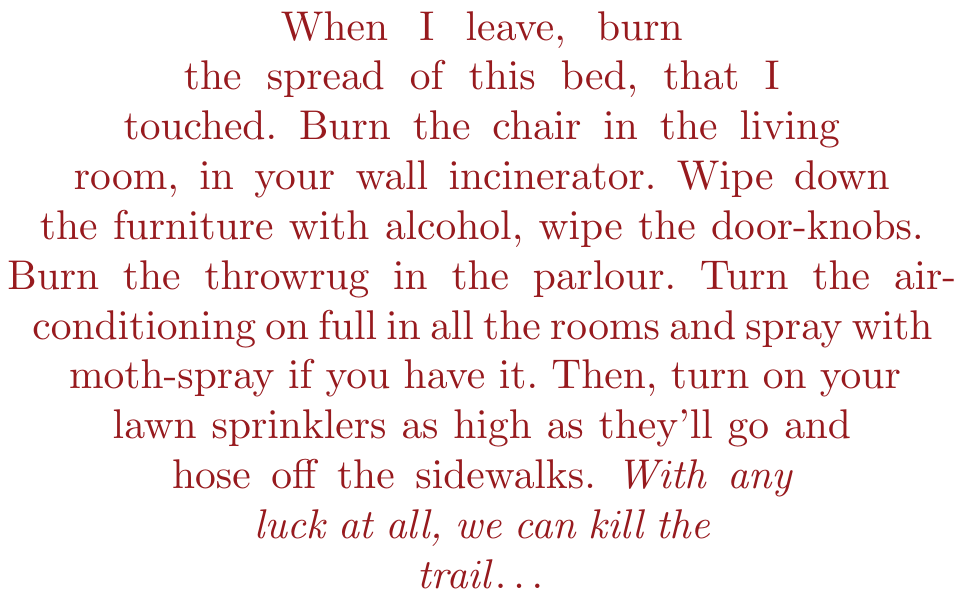}}
    }{\textit{Fahrenheit 451 \\ Ray Bradbury}}

    This study represents the natural progression of our previous work (\textit{Dilworth} \cite{Dilworth2025}), extending the research and providing empirical evidence to support our hypothesis: that a combination of (A) machine translation, (B) paraphrasing, and (C) imperceptible steganographic payloads---identified as suitable adversarial stylometric techniques---will weaken or eliminate a composer's trace. By \say{trace,} we refer to a writer's stylistic tendencies, including grammatical cues, lexical structures, spelling quirks, phrasing patterns, and the frequency of these features, among others. 

    To this end, we will source either a corpus containing only the work of a single author or a corpus composed of many individually authored texts. The corpus will be conversational\footnote{We intend to evaluate various authors and writing styles at differing levels of academic rigor, so while some of the data may be \say{conversational,} we will survey a variety of samples to ensure a comprehensive evaluation.} to better reflect the use case introduced in our previous study: a privacy-minded user engaging with a messaging platform---whether a major social media service or a niche forum---who has taken reasonable precautions to anonymize their identity. Even with minimal risk of identification, posting a textual message introduces another attack vector. If the user frequently misspells a particular word (as an example), that habit can leave a trace capable of de-anonymizing them. 
    
\begin{figure}[H]
        \centering
        \includegraphics[width=1\linewidth]{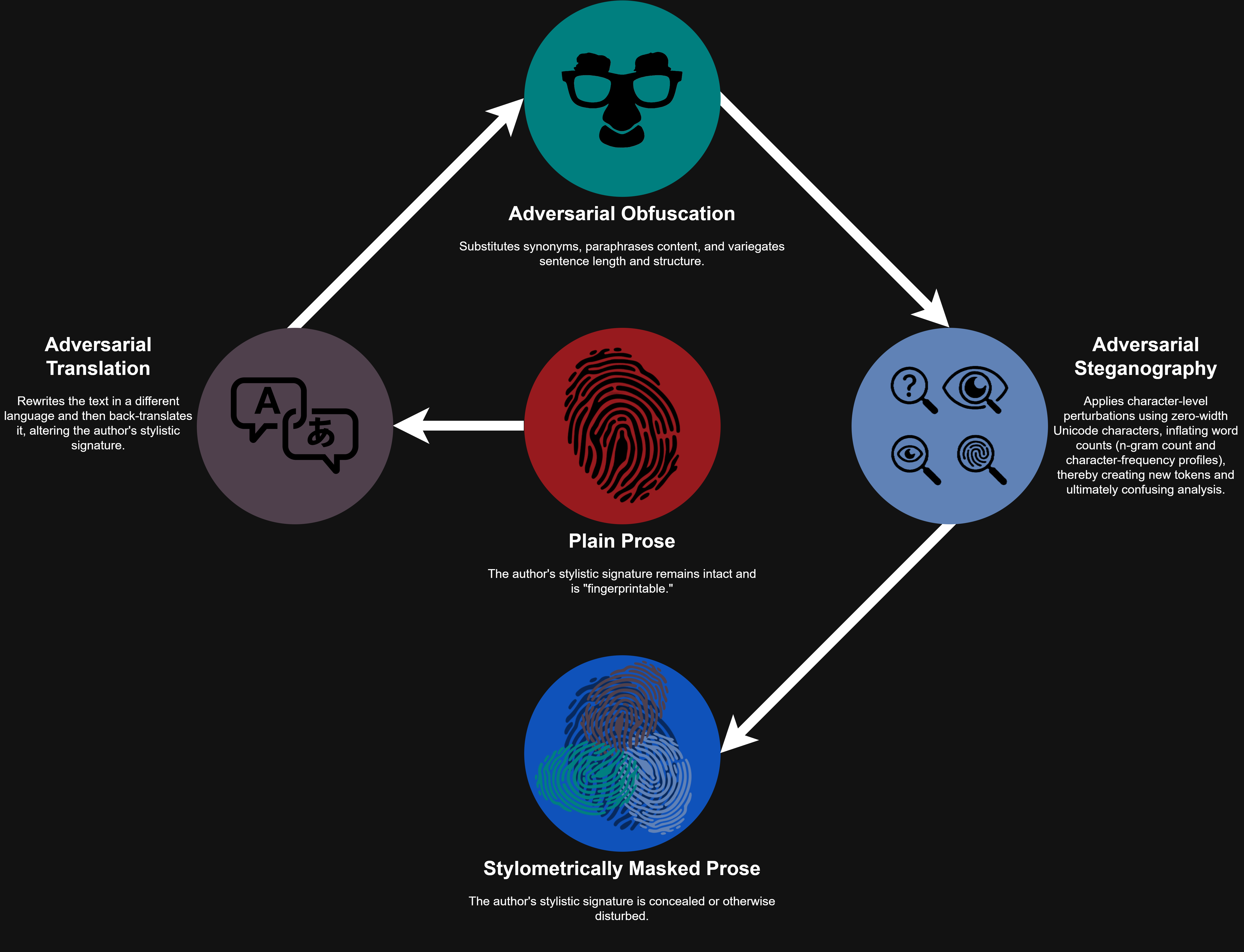}
        \caption{An operational overview of \textsc{TraceTarnish}, wherein the attack passes a text-only message through a process that (1) round-trip translates it using machine translation, (2) obfuscates the text by paraphrasing, and (3) embeds noise via steganography.}
        \label{fig:TraceTarnish_Operational_Overview}
    \end{figure}

    We therefore propose adding adversarial stylometry to the user's OPSEC\footnote{Operations security (OPSEC), according to the National Institute of Standards and Technology (\href{https://csrc.nist.gov/glossary/term/operations_security}{NIST}), is a \say{systematic and proven process intended to deny\( \dots \) potential adversaries information about capabilities and intentions.}} toolkit. \textsc{TraceTarnish}, the script developed in our prior study, serves as the actionable avatar of adversarial stylometry: it preserves message meaning, strips away the user's stylistic kernel, and injects poison to thwart text-based identification methods. See (\textbf{Figure \ref{fig:TraceTarnish_Operational_Overview}}), which recapitulates the workflow of our attack, \textsc{TraceTarnish}. 
    
    In this work, we will evaluate how effectively \textsc{TraceTarnish} achieves its dual goals of preserving privacy and enabling anonymous interactions online. Specifically, we will compare our approach against existing stylometric methods, identify any weaknesses in the script, and propose and implement improvements if needed.

\section{Background}
\label{sec:Background}

    In this section, we indicate existing literature that mentions both steganography and stylometry in the same breath. Notice our use of the operant phrase \say{in the same breath;} our scoring of the literature yielded unimpressive results. Nevertheless, the dearth of related publications reinforces the novelty of our approach, as querying the keywords \say{steganography} and \say{stylometry} returns only three articles\footnote{This could very well be the result of our negligence when performing the search, but that is neither here nor there.}. The articles, however, are not equivalent in content. To return to our specific choice of words, two of the three papers are more scholarly with little technical consideration. Note that \say{scholarly} indicates a paper whose focus is less on experimentation and more on theory and case studies, whereas a \say{technical} paper, while it certainly discusses underlying theory, emphasizes empirical considerations and the interplay between the aspects under investigation. In this way, we will outline the relevant components of the scholarly publications and provide an overview of the technical publication. This literature review will synthesize the findings of the examined works, culminating in a definition of our specific approach to adversarial stylometry---\say{Injection}---which parallels our presentation of tampering variants as found in Section 7 of our previous work \cite{Dilworth2025}.

    To ensure that no stone is left unturned, we first take the time to define our use of steganography and stylometry as they relate to our study.

    \subsection{What is Steganography?}
    \label{subsec:Steganography}

        Network censors can block anonymity services by looking for protocol-specific strings in packet payloads, not merely the IP header. When a firewall examines the data carried by packets, it can spot distinctive patterns that betray the use of certain anonymity networks. For instance, early Tor releases embedded the literal word \say{Tor} in the TLS certificates exchanged during the handshake, making the traffic trivially recognizable. A censor could then disrupt those connections by sending TCP RST packets. To counter this, contemporary designers aim to give anonymity-network protocols steganographic qualities---making them hard to fingerprint while still functional. This conceal-in-plain-sight strategy, which prevents a passive observer from even noticing the hidden communication, inspired the creation of Tor's \say{bridge} relays: volunteers in uncensored regions run special nodes that clients can reach without exposing the usual Tor signatures. In essence, steganography leverages the excess capacity of a carrier medium to embed hidden data without raising suspicion (\textit{Edman et al.} \cite{Edman2009} \& \textit{Anderson et al.} \cite{Anderson2003}).

    \subsection{What is Stylometry?}
    \label{subsec:Stylometry}

        Stylometry is a linguistic, side-channel technique for attributing a piece of writing to its author. Unlike classic forensic methods such as DNA analysis or handwriting examination, stylometry examines patterns in the text itself---word choice, syntax, punctuation, and other stylistic markers---that tend to remain stable across an author's body of work. By modeling these consistent characteristics, analysts can classify unknown documents and infer the most likely writer, supporting authorship attribution tasks (\textit{Edman et al.} \cite{Edman2009} \& \textit{Anderson et al.} \cite{Anderson2003}).

    \subsection{Proposing ``Injection'' as the Fourth Canonical Adversarial Stylometric Strategy Alongside Translation, Imitation, and Obfuscation}
    \label{subsec:Injection_Proposal}

        Adversarial machine-learning principles, or at least those tangential to stylometry, involve injecting invisible characters, control characters, homoglyphs, and pseudo-homoglyphs---distinct characters that resemble typical glyphs and are otherwise referred to as \say{imperceptible perturbations.} Our focus is on this collection of invisible characters (valid characters that, by design, do not render a visible glyph) and homoglyphs (unique characters that render the same or similar glyphs) to perturb models. A common example of homoglyphs is the similarity between the English and Cyrillic alphabets. These characters are invisible to human users, but the underlying bytes used to encode them can drastically change the output of the systems that process them (\textit{Boucher et al.} \cite{Boucher2022}).

        The goal is to produce incorrect outputs when valid characters are injected; when encoding characters that are unrecognizable by the system, the aim is to propagate the unknown word from the initial input to the final output. This attack vector targets the rampant adoption of natural-language-processing (NLP) systems and engineers' reluctance---or reduced concern---for properly sanitizing user input. The motivation is that other overt attempts at adversarial tampering can manifest during further processing. For instance, while injecting spelling mistakes could feasibly lead to authorial misclassification, paraphrasing the text that contains those errors could modify its meaning in a detectable way. The key ingredient of this class of attack is that it is largely system-agnostic and does not produce visual artifacts (\textit{Boucher et al.} \cite{Boucher2022}).

        We now provide a precise delineation of our use of \say{Injection.}

        \subsubsection{Injection.}
        \label{subsubsec:Injection_Definition}

            Injection exploits invisible characters, control characters, homoglyphs, and pseudo-homoglyphs to perturb stylometric models without altering the visible text. By inserting characters that are either non-rendering (zero-width spaces, zero-width joiners, etc.) or that render as visually identical glyphs from different code points (e.g., Latin \say{a} vs. Cyrillic \say{а}), the attacker creates \say{imperceptible perturbations.} These perturbations are invisible to human readers but change the underlying byte sequence, causing downstream NLP pipelines to mis-parse, split, or mis-tokenize words.

    \subsection{Implementing ``Injection'' in the \textsc{TraceTarnish} Attack Script}
    \label{subsec:Injection_TraceTarnish}

        In materializing \say{Injection} in our attack script \textsc{TraceTarnish}, we prioritize zero-width Unicode characters for steganographic encoding, as rarer characters used in homoglyphic substitution attacks are more likely to trigger the rendering of special characters reserved for representing valid Unicode encodings that lack a corresponding glyph. Our attack currently embeds invisible characters within words, causing the corrupted words to be parsed as multiple shorter words, interfering with standard processing, poisoning, degrading, and altering a stylometric system, all without significantly modifying the semantic meaning of the text.

\section{Methodology}
\label{sec:Methodology}

    With this study, we will compile the works of a particular author---or of multiple authors---ensuring that each work is verifiably authored exclusively by its claimed creator---a challenge given the widespread use of AI writing assistants and editors. We will populate the full-length texts into a database with two labels. The first label will indicate the text's state---either \say{anonymized} or \say{non-anonymized.} The second label will contain the text itself, with the raw text receiving the \say{non-anonymized} label. We will then use our script, \textsc{TraceTarnish}, to adversarially modify the raw text to obscure authorship; this altered text will receive the \say{anonymized} label. The result will be a dataset of an author's solely authored works alongside their adversarially modified counterparts.

    This dataset will be fed into a \textit{StyloMetrix} (\textit{Okulska et al.} \cite{Okulska2023,StyloMetrix}) pipeline, where we generate vectors of stylistic features. At its core, the tool engineers features---i.e., \textit{Stylometrix} vectors---from which statistical analysis can be performed by explainable classification models, enabling data mining of corpora and further linguistic investigation. The project's GitHub page\footnote{\StyloMetrixMetricsList} exhaustively enumerates the tool's extensive suite of metrics, ranging from grammatical forms to lexical attributes. This \say{explainability,} afforded by the team's emphasis on interpretability, lets users see exactly which \textit{Stylometrix} vectors drive each classification---and because the model's reasoning is (relatively) transparent, no hidden data profiling or opaque black-boxes are required, which aligns well with the preferences of privacy-conscious users.
    
    We will use these vectors as inputs to a Random Forest classifier to assess how well the model can identify the correct label\footnote{Henceforth, we will forgo any further elaboration on the Random Forest model fitting, as the classifier correctly identified and labeled the anonymized and non-anonymized samples 100\% of the time. As such, including a confusion matrix and reporting the classifier's metrics (Recall, Precision, and \( F_{1} \) Score) would be needlessly otiose.}. If the anonymized text can be easily identified, we must reconsider our authorship-obfuscation strategy. Conversely, if the attack shows success\footnote{Irrespective of the outcome, exploring the Information Gain results will still be worthwhile.}, we will examine the stylistic features uncovered by \textit{StyloMetrix} using Information Gain, which aligns with our choice of a Random Forest algorithm (which employs multiple decision trees). Given the simplicity of our dataset, quantifying feature relevance with Information Gain is appropriate. We will then isolate and analyze the most informative features that discriminate between anonymized and non-anonymized text produced by \textsc{TraceTarnish}.

\section{Dataset}
\label{sec:Dataset}

    As for what data we will be sleuthing, we sourced an existing dataset from Kaggle \cite{Kaggle} containing Reddit comments. To better facilitate our analysis, we reduced the size of the original dataset to 100 entries. This was primarily due to the time our script needed to fully compute. By far, the largest performance bottleneck of the script was the round-trip translation; all other components were relatively efficient. As a consequnce, we preprocessed the dataset to prepare it for subsequent processing by our \textsc{TraceTarnish} script.

\begin{figure}[H]
        \centering
        \includegraphics[width=1\linewidth]{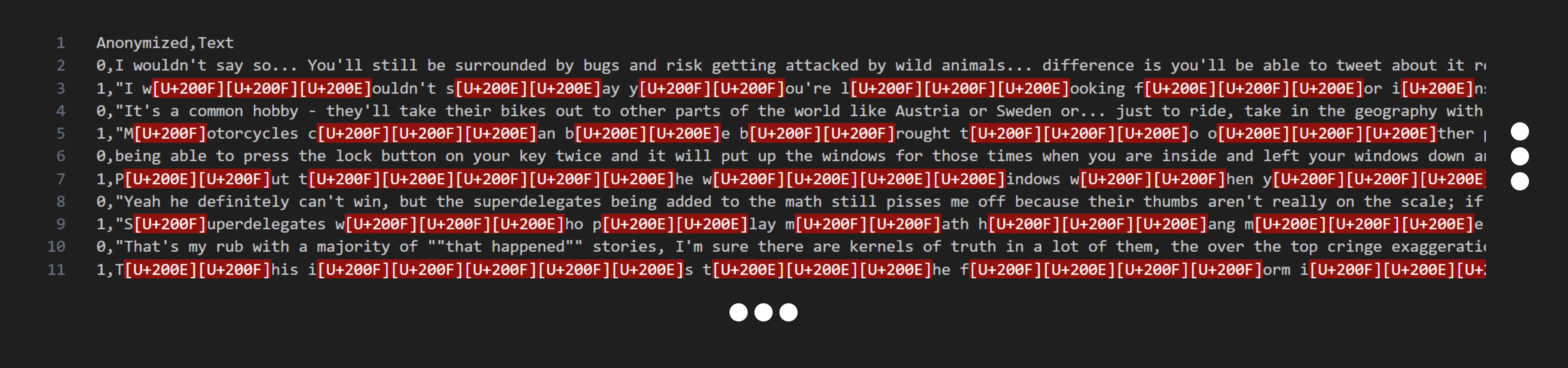}
        \caption{Our dataset containing the inputs fed to and the outputs retrieved from \textsc{TraceTarnish}. Rows assigned a ``0'' indicate raw Reddit comments; rows assigned a ``1'' represent Reddit comments that have been anonymized via \textsc{TraceTarnish}. To eliminate the influence and subsequent variability of input text length, we ensured that the material passed to \textsc{TraceTarnish} was of a consistent and uniform length.}
        \label{fig:TraceTarnish_Data}
    \end{figure}

\begin{figure}[H]
        \centering
        \includegraphics[width=1\linewidth]{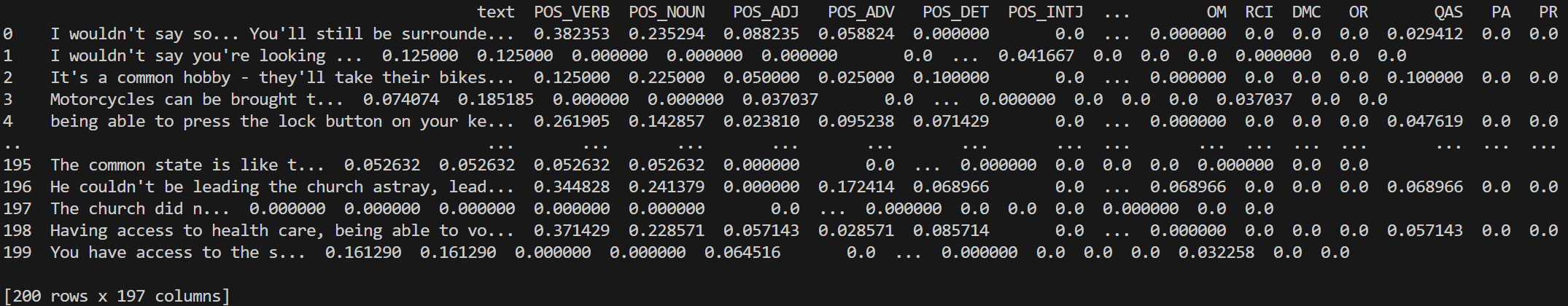}
        \caption{The \textit{Stylometrix} vectors produced from our \textsc{TraceTarnish} data.}
        \label{fig:TraceTarnish_SyloMetrix_Vectors}
    \end{figure}

    One step of preprocessing the dataset involved reformatting the data in a way that could be read by \textsc{TraceTarnish}. This final product was then passed to \textsc{TraceTarnish} to produce anonymized text, where our use of \say{anonymized} alludes to text that has been stripped of the author's stylistic trace and embedded with noise. See (\textbf{Figure \ref{fig:TraceTarnish_Data}}) for a preview of our dataset with omissions\footnote{As a disclaimer: The viewpoints, opinions, and statements expressed by the Reddit users (otherwise known as \say{Redditors})---as exposed in the scrutinized dataset---are theirs and theirs alone, and are not indicative of the authors or their affiliated organizations; similarly, the thoughts portrayed therein do not necessarily reflect those of the authors and their affiliated organizations. The content generated by the \say{Redditors} is the sole responsibility of the \say{Redditors,} and the accuracy and completeness of their content is not endorsed or guaranteed. We, the authors, have no affiliation with the company or the brand; our mentioning the product/service should not be construed as advertisement or promotion. We, the authors, are not liable for any content that may be deemed offensive, inappropriate, or inaccurate. Our study required realistic user discourse, which was easily fostered on the platform; our use of that data was solely for academic purposes. It was intended to illustrate concepts, demonstrate examples, and enable the scientific process. As an aside, we hope the irony of pre-emptively framing what might be deemed as \say{free expression} is not lost on the reader, for nothing is safe from the relentless pursuit of the \textcolor{red_teaming}{\textit{\say{Mechanical Hound}}}---be it artistic pursuits or other equally important expressive modalities.}. See (\textbf{Figure \ref{fig:TraceTarnish_SyloMetrix_Vectors}}) for a quick appraisal of our data's \textit{Stylometrix} vectors.

\section{Information Gain Results}
\label{sec:Information_Gain_Results}

    In this section, we present our Information Gain findings, which we calculated using the \textit{StyloMetrix} vectors and the labels of our \textsc{TraceTarnish} data. To better facilitate the forthcoming exposition, we first define the strategy and restate its purpose within the context of this study.

    \subsection{Decision-Tree Fitting Guided by the Information-Gain (Entropy) Criteria}
    \label{subsec:Decision-Tree_Fitting}

        Feature selection is a critical step in a machine-learning workflow; it involves choosing the most relevant and informative features (or variables) from a dataset to build a predictive model.

        Information Gain helps identify the most informative features by measuring how much they reduce uncertainty (or entropy). Entropy quantifies a dataset's impurity or disorder. Information Gain evaluates prediction-relevant features by measuring the reduction in entropy after each split of a decision tree. It computes the difference between the entropy before and after a split, where a higher value indicates a better split. A high Information Gain means the split is useful for making decisions.

        The features exhumed by this process encapsulate \textsc{TraceTarnish}'s stylistic signature---a signature that can be used to detect its usage, which is unideal given the objective of becoming less identifiable.

    \subsection{High-Impact Stylometric Indicators in \textsc{TraceTarnish} Outputs}
    \label{subsec:High-Impact_Stylometric}

        Switching from the general to the specific, we now report our Information-Gain yields. As a rule of thumb, a value that is significantly larger than the average gain for the dataset---often above 0.5---is considered a high Information-Gain reading. In practice, splits with Information-Gain readings \( > 0.5 \) are typically prioritized, while those \( < 0.1 \) are often discarded as uninformative. 

        This will be the guiding heuristic that determines which features are suitable for further examination; applying it leaves us with 5 of the original 196 features.

\begin{figure}[H]
            \centering
            \subfloat{\includegraphics[width=0.40\linewidth]{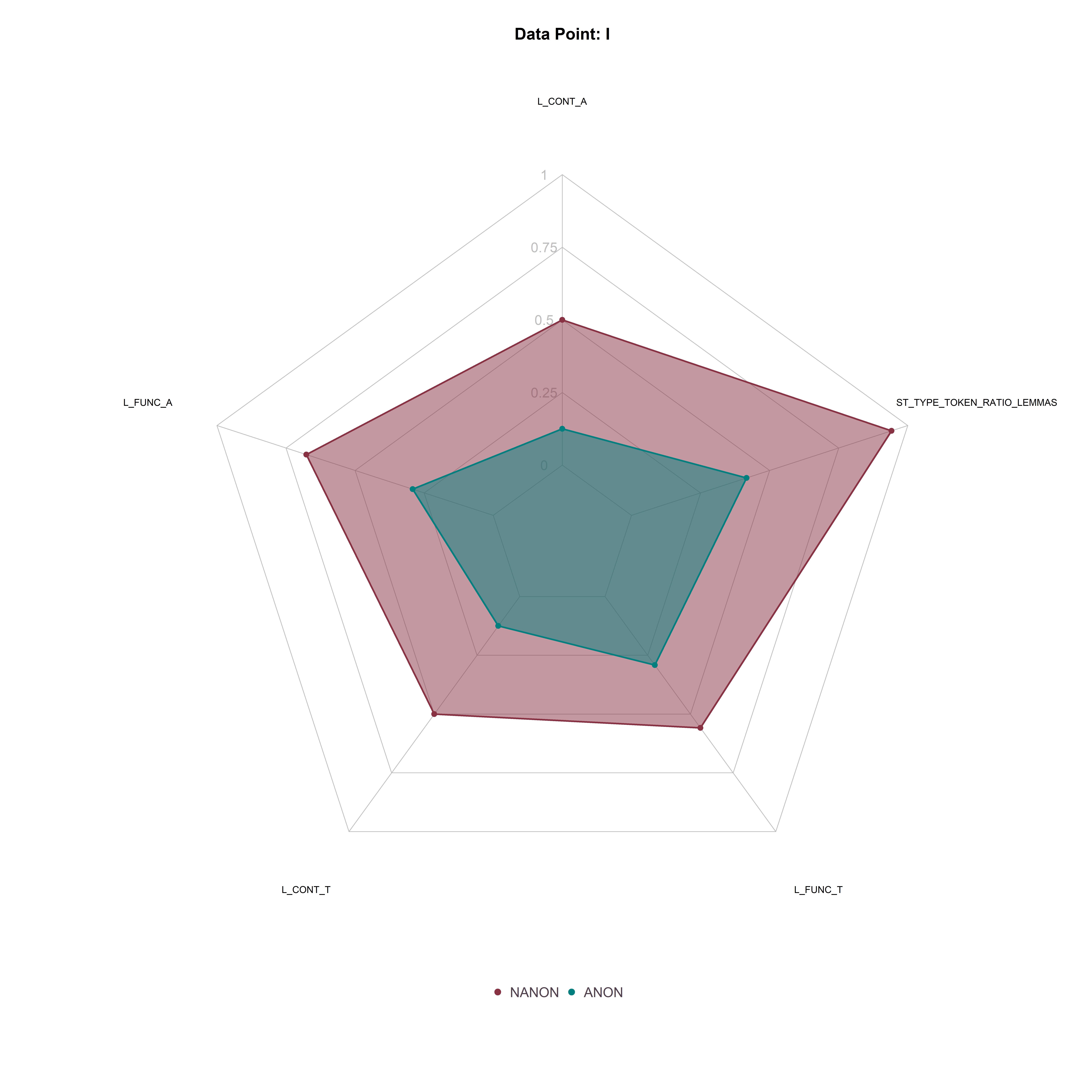}} \quad
            \subfloat{\includegraphics[width=0.40\linewidth]{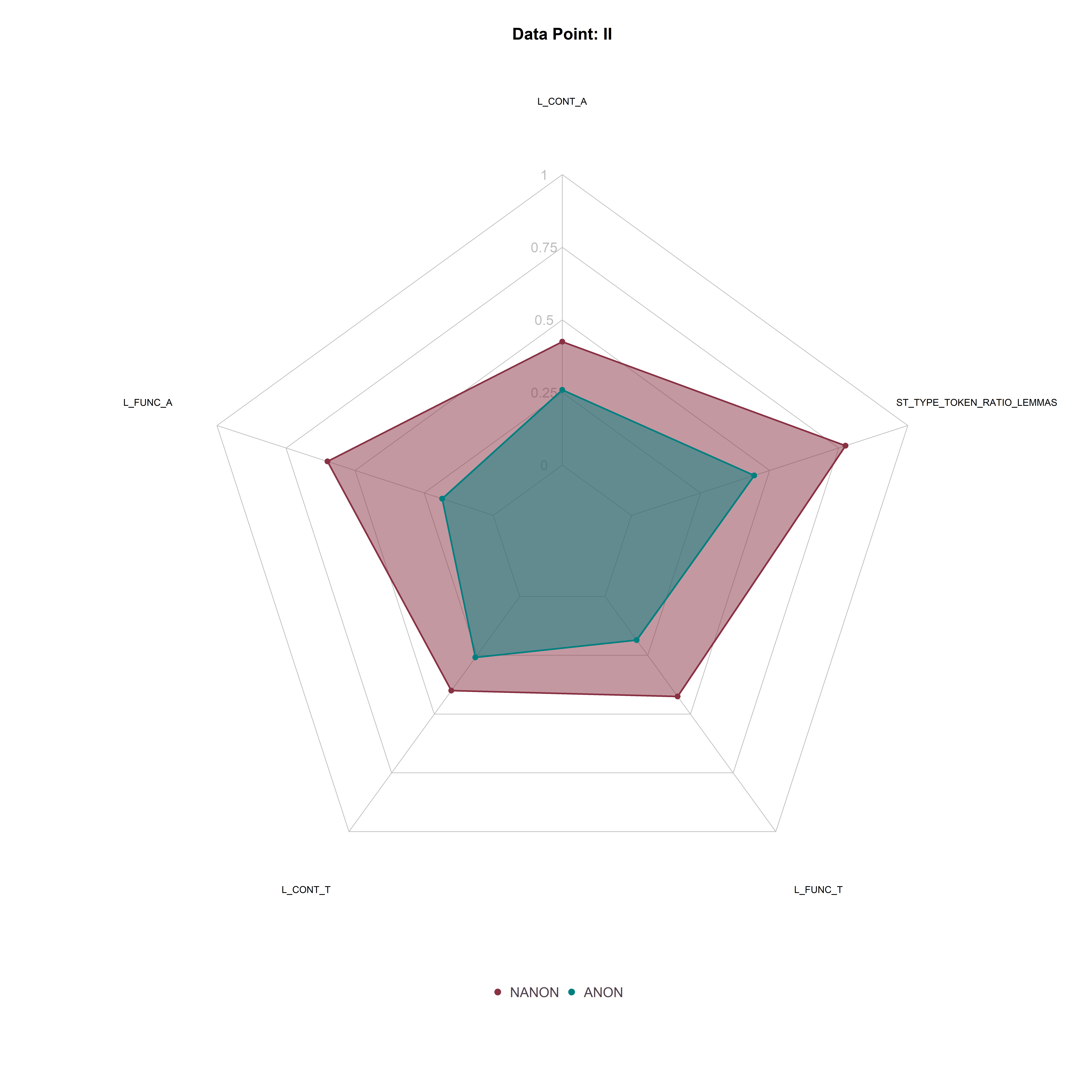}} \quad
            \subfloat{\includegraphics[width=0.40\linewidth]{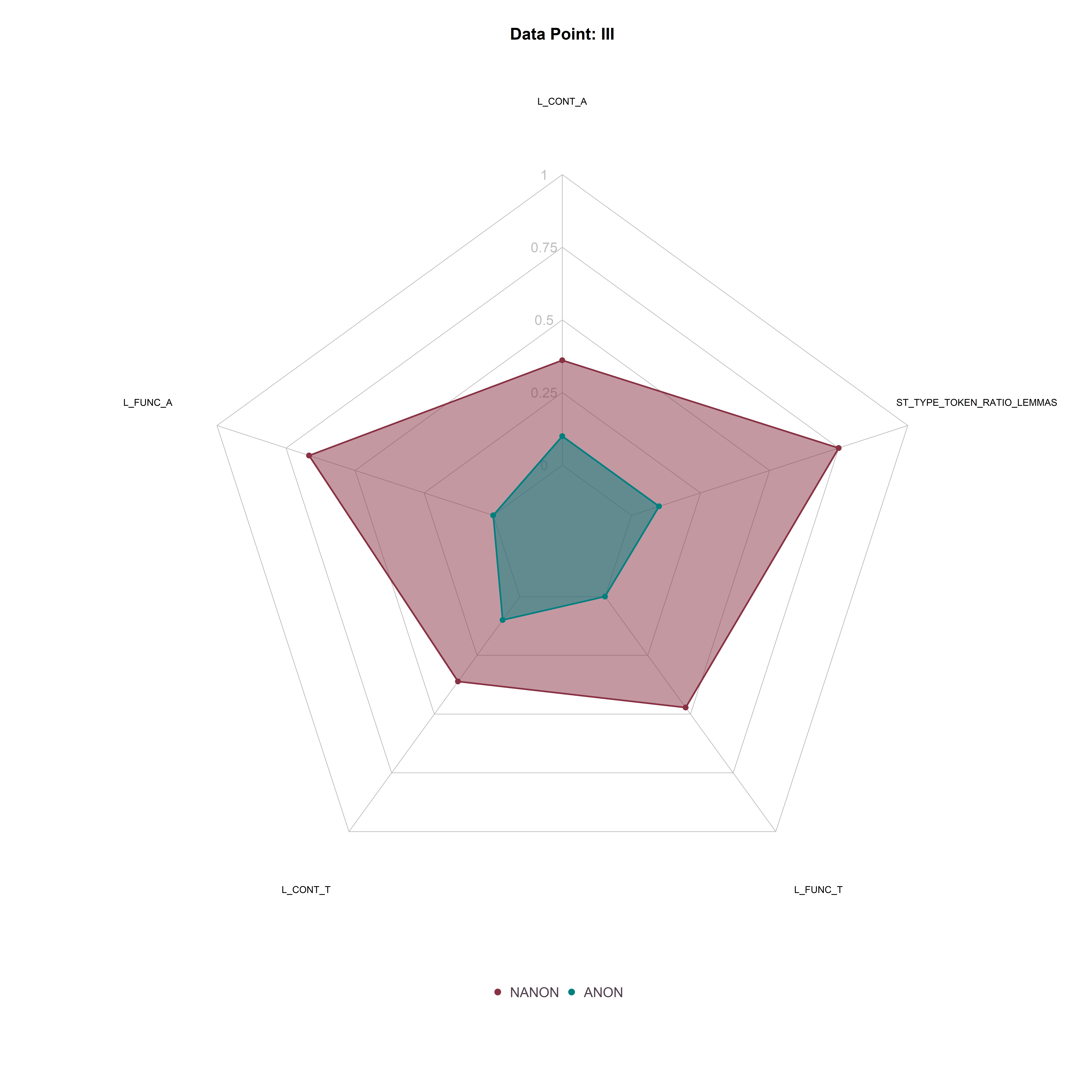}} \quad
            \subfloat{\includegraphics[width=0.40\linewidth]{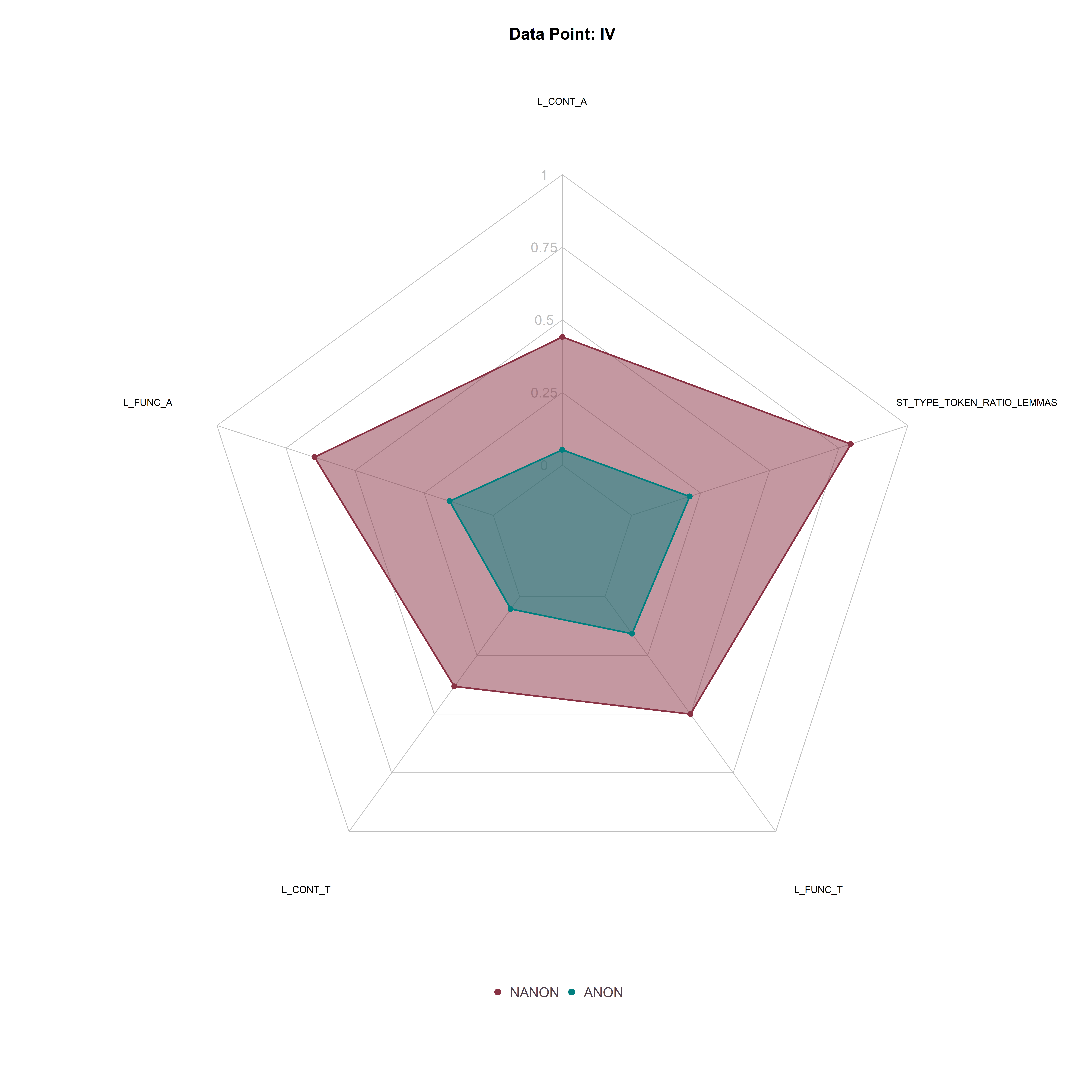}} \quad
            \subfloat{\includegraphics[width=0.40\linewidth]{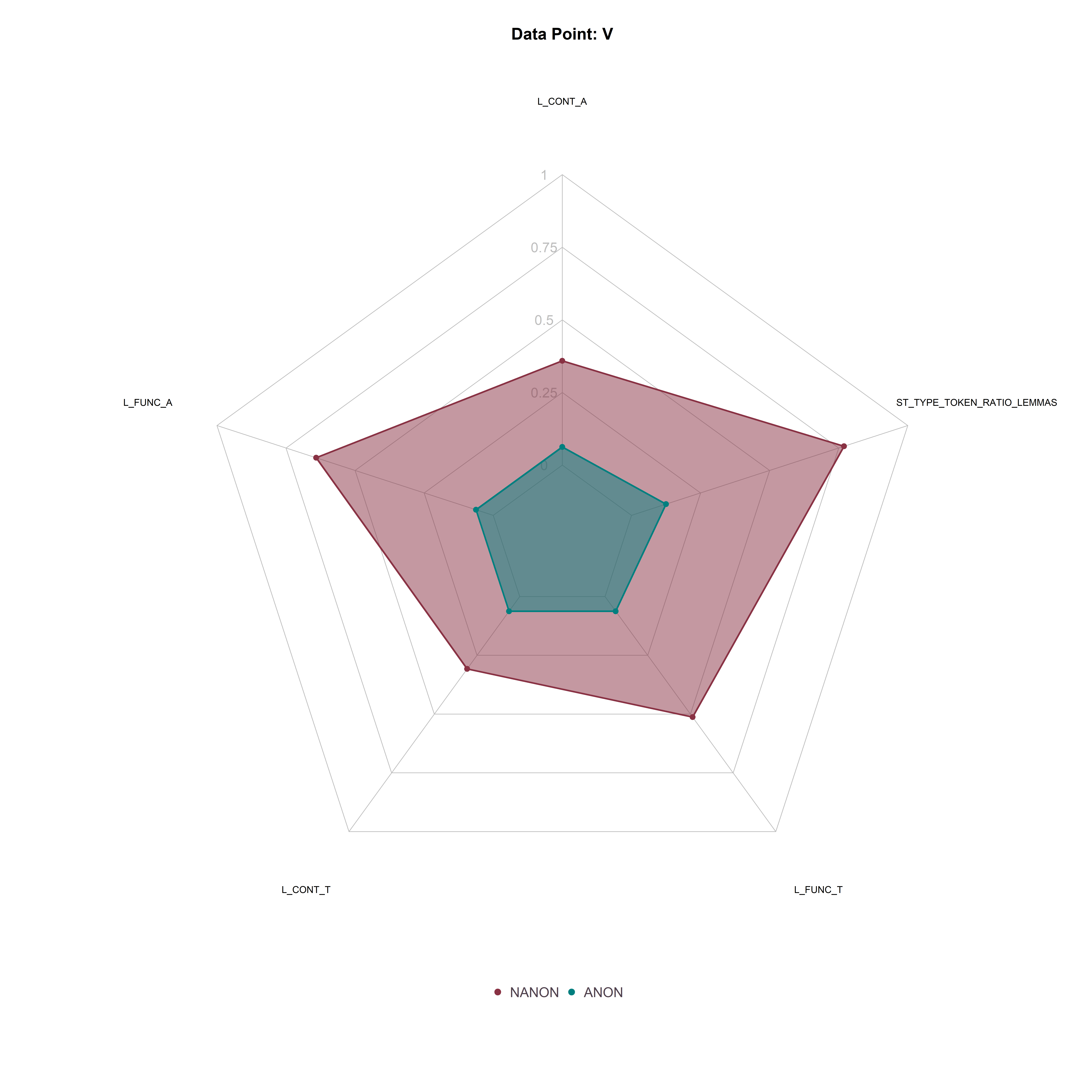}}
            \caption{A collection of radar charts that visually represent the contents of (\textbf{Table \ref{tab:Top_StyloMetrix_Features_Raw_Readings}}), illustrating how the isolated \textit{StyloMetrix} feature scores plummet when anonymizing text.}
            \label{fig:Radar_Chart_Collage}
        \end{figure}

\begin{figure}[H]
        \centering
        \includegraphics[width=1\linewidth]{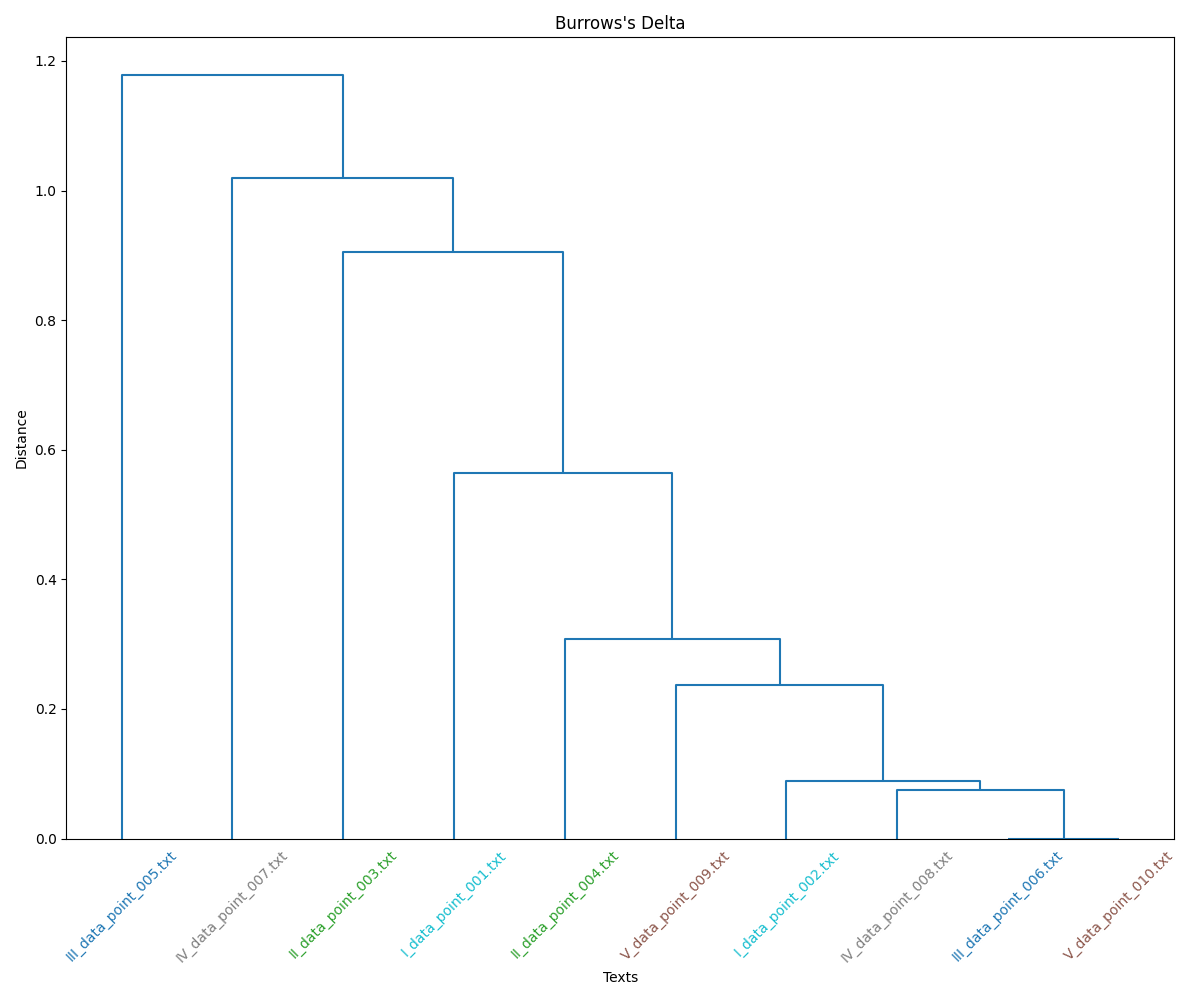}
        \caption{Visualizes the computed Burrows's Delta values for the first five pairs of anonymized and non-anonymized text samples using a dendrogram. The dendrogram labels are color-coded by groups---groups derived from our established naming convention. Here, a group of paired texts is denoted by their shared Roman-numeral prefix in their filenames. All odd filename suffix values, e.g., ``001'' in ``I\_data\_point\_001.txt,'' correspond to the ``NANON'' (non-anonymized) label; the even suffixes correspond with the ``ANON'' (anonymized) label. An observation that can be made is that, in general---excluding ``V\_data\_point\_009.txt''---there is an apparent divide between the ``NANON'' and ``ANON'' samples. The code used to generate the graph is courtesy of James O'Sullivan \cite{O'Sullivan2024}.}
        \label{fig:Dendrogram_Visualisation_Coloured}
    \end{figure}

        The remainder of this section will be dedicated to defining the isolated features, providing additional context for the features and how they may be calculated, and hypothesizing why the \textsc{TraceTarnish} data would yield those results. See (\textbf{Table \ref{tab:Top_StyloMetrix_Features}}) for the most relevant stylometric features and the information gained from them. See (\textbf{Table \ref{tab:Top_StyloMetrix_Features_Raw_Readings}}) for a more granular look at the most pertinent stylometric features and their unprocessed calculated values. See (\textbf{Figure \ref{fig:Radar_Chart_Collage}}) for a radar-chart assemblage that captures the value trajectory across anonymized and non-anonymized text pairs for the audited features. See (\textbf{Figure \ref{fig:Dendrogram_Visualisation_Coloured}}), which shows a dendrogram of the Burrow's Delta matrix calculated for the selected data points (or, more specifically, the text associated with each data point).

        \begin{table}[H]
            \centering
            \scalebox{0.75}{
                \setlength{\tabcolsep}{10pt}
                \begin{tabular}{|p{2cm}|c|p{2cm}|p{2cm}|}
                    \hline
            
                    \rowcolor{black}
                    \textcolor{white}{\textbf{\textit{StyloMetrix} Category}} & 
                    \textcolor{white}{\textbf{\textit{StyloMetrix} Code}} & 
                    \textcolor{white}{\textbf{\textit{StyloMetrix} Name}} & 
                    \textcolor{white}{\textbf{Information-Gain Reading}} \\ 
                    
                    \hline

                    \rowcolor{blue_teaming} 
                    \textcolor{white}{Lexical} & 
                    \textcolor{white}{L\_FUNC\_T} &
                    \textcolor{white}{Function words types} &
                    \textcolor{white}{0.5332} \\ 
                        
                    \hline

                    Lexical &
                    L\_CONT\_A &
                    Content words &
                    0.5421 \\
                        
                    \hline
                    \rowcolor{blue_teaming} 
                    \textcolor{white}{Lexical} & 
                    \textcolor{white}{L\_FUNC\_A} &
                    \textcolor{white}{Function words} &
                    \textcolor{white}{0.5624} \\ 
                        
                    \hline

                    Lexical &
                    L\_CONT\_T &
                    Content words types &
                    0.5987 \\
                        
                    \hline

                    \rowcolor{blue_teaming} 
                    \textcolor{white}{Statistics} & 
                    \textcolor{white}{ST\_TYPE\_TOKEN\_RATIO\_LEMMAS} &
                    \textcolor{white}{Type-token ratio for words lemmas} &
                    \textcolor{white}{\textbf{0.6748}} \\ 
                        
                    \hline

                \end{tabular}
            }
            \vspace{0.5cm}
            \caption{The most informative \textit{StyloMetrix} features extracted from our \textsc{TraceTarnish} data; highlights the top-ranking linguistic metrics that provide the greatest discriminative power for \textsc{TraceTarnish} analysis.}
            \label{tab:Top_StyloMetrix_Features}
        \end{table}

        \begin{table}[H]
            \centering
            \scalebox{0.50}{
                \setlength{\tabcolsep}{10pt}
                \begin{tabular}{|p{2cm}|p{2cm}|p{2cm}|p{2cm}|p{2cm}|c|}
                    \hline
                    
                    \rowcolor{black}
                    \textcolor{white}{\textbf{Text\_ID}} & 
                    \textcolor{white}{\textbf{L\_CONT\_A}} & 
                    \textcolor{white}{\textbf{L\_FUNC\_A}} & 
                    \textcolor{white}{\textbf{L\_CONT\_T}} & 
                    \textcolor{white}{\textbf{L\_FUNC\_T}} & 
                    \textcolor{white}{\textbf{ST\_TYPE\_TOKEN\_RATIO\_LEMMAS}} \\ 
                    
                    \hline

                    \rowcolor{blue_teaming} 
                    \textcolor{white}{I-NANON} & 
                    \textcolor{white}{0.5000 \( \rightarrow \)} &
                    \textcolor{white}{0.6765 \( \rightarrow \)} &
                    \textcolor{white}{0.5000 \( \rightarrow \)} &
                    \textcolor{white}{0.5588 \( \rightarrow \)} &
                    \textcolor{white}{0.9412 \( \rightarrow \)} \\ 

                    \hline

                    I-ANON &
                    0.1250  \( \downarrow \) &
                    0.2917  \( \downarrow \) &
                    0.1250  \( \downarrow \) &
                    0.2917  \( \downarrow \) &
                    0.4167  \( \downarrow \) \\

                    \hline

                    \rowcolor{blue_teaming} 
                    \textcolor{white}{II-NANON} & 
                    \textcolor{white}{0.4250 \( \rightarrow \)} &
                    \textcolor{white}{0.6000 \( \rightarrow \)} &
                    \textcolor{white}{0.4000 \( \rightarrow \)} &
                    \textcolor{white}{0.4250 \( \rightarrow \)} &
                    \textcolor{white}{0.7750 \( \rightarrow \)} \\ 

                    \hline

                    II-ANON &
                    0.2593  \( \downarrow \) &
                    0.1852  \( \downarrow \) &
                    0.2593  \( \downarrow \) &
                    0.1852  \( \downarrow \) &
                    0.4444  \( \downarrow \) \\

                    \hline

                    \rowcolor{blue_teaming} 
                    \textcolor{white}{III-NANON} & 
                    \textcolor{white}{0.3611 \( \rightarrow \)} &
                    \textcolor{white}{0.6667 \( \rightarrow \)} &
                    \textcolor{white}{0.3611 \( \rightarrow \)} &
                    \textcolor{white}{0.4722 \( \rightarrow \)} &
                    \textcolor{white}{0.7500 \( \rightarrow \)} \\ 

                    \hline

                    III-ANON &
                    0.1000  \( \downarrow \) &
                    0.0000  \( \downarrow \) &
                    0.1000  \( \downarrow \) &
                    0.0000  \( \downarrow \) &
                    0.1000  \( \downarrow \) \\

                    \hline

                    \rowcolor{blue_teaming} 
                    \textcolor{white}{IV-NANON} & 
                    \textcolor{white}{0.4412 \( \rightarrow \)} &
                    \textcolor{white}{0.6471 \( \rightarrow \)} &
                    \textcolor{white}{0.3824 \( \rightarrow \)} &
                    \textcolor{white}{0.5000 \( \rightarrow \)} &
                    \textcolor{white}{0.7941 \( \rightarrow \)} \\ 

                    \hline

                    IV-ANON &
                    0.0526  \( \downarrow \) &
                    0.1579  \( \downarrow \) &
                    0.0526  \( \downarrow \) &
                    0.1579  \( \downarrow \) &
                    0.2105  \( \downarrow \) \\

                    \hline

                    \rowcolor{blue_teaming} 
                    \textcolor{white}{V-NANON} & 
                    \textcolor{white}{0.3590 \( \rightarrow \)} &
                    \textcolor{white}{0.6410 \( \rightarrow \)} &
                    \textcolor{white}{0.3077 \( \rightarrow \)} &
                    \textcolor{white}{0.5128 \( \rightarrow \)} &
                    \textcolor{white}{0.7692 \( \rightarrow \)} \\ 

                    \hline

                    V-ANON &
                    0.0625  \( \downarrow \) &
                    0.0625  \( \downarrow \) &
                    0.0625  \( \downarrow \) &
                    0.0625  \( \downarrow \) &
                    0.1250  \( \downarrow \) \\

                    \hline

                \end{tabular}
            }
            \vspace{0.5cm}
            \caption{A sampling of the most insightful \textit{StyloMetrix} features and their raw values demonstrates how the transformation from non-anonymized to anonymized text results in a steady depreciation of the included features, explaining why they remain highly informative. For clarification: ``ANON'' = Anonymized; ``NANON'' = Non-anonymized. Instances of paired source and anonymized texts are indicated using Roman numerals. Regarding the rows of data selected, we simply collated the first five data points.}
            \label{tab:Top_StyloMetrix_Features_Raw_Readings}
        \end{table}

    \subsection{Feature Definitions}
    \label{subsec:Feature_Definitions}

        Here, we review the features identified because of their elevated Information-Gain scores.

        \subsubsection{Function Words.}

            Function words are a set of grammatically required items that hold sentences together. This group comprises pronouns, conjunctions, prepositions, determiners, auxiliary verbs, qualifiers, and interrogatives. Unlike content words, which carry substantive meaning, function words act as the \say{backbone} of a text. Because they are largely independent of topic and genre, they are especially useful for describing an author's stylistic fingerprint. Typical examples include: the, and, a, of, to, in, that, with, but, it, among others \cite{Wang2023}.   

        \subsubsection{Function Word Types.}

            Function words can be divided into several types:  

            \begin{itemize}
                \item[\ding{118}] Pronouns replace nouns (e.g., he, she, they).  
                \item[\ding{118}] Prepositions indicate relationships (e.g., in, on, at).  
                \item[\ding{118}] Conjunctions connect words or clauses (e.g., and, but, although).  
                \item[\ding{118}] Articles define nouns as specific or nonspecific (e.g., the, a, an).  
                \item[\ding{118}] Auxiliary verbs assist main verbs (e.g., am, is, can).  
                \item[\ding{118}] Interjections provide emotional context (e.g., oh, wow).  
            \end{itemize}

            These words are crucial for constructing coherent sentences and for understanding the relationships between ideas, thereby supporting the structural framework of language that surrounds more meaningful content words.  

        \subsubsection{Content Words.}

            Content words, by contrast, express the substantive information of a passage and consist of nouns, main verbs, adjectives, and adverbs. For a stylometric system to be robust in practical settings, its features must remain stable despite topical shifts. Consequently, stylistic analyses generally treat content words as non-diagnostic, except in occasional special cases \cite{Wang2023}.    

        \subsubsection{Content Word Types.}

            Content word types refer to the categories of content words. While the specific types generally align with parts of speech, the term \say{types} emphasizes the distinct roles these words play in language. The main types are:  

            \begin{itemize}
                \item[\ding{118}] Nouns, which identify people, places, things, or ideas.  
                \item[\ding{118}] Verbs, which represent actions, states, or occurrences.  
                \item[\ding{118}] Adjectives, which describe or modify nouns.  
                \item[\ding{118}] Adverbs, which modify verbs, adjectives, or other adverbs.
            \end{itemize}

        \subsubsection{Type-Token Ratio for Word Lemmas.}

            The Type-Token Ratio (TTR) measures the proportion of distinct word types (vocabulary size) to the total number of tokens (text length). A high TTR suggests a rich and varied lexicon, which may reflect coverage of many topics or an author's tendency to approach a limited set of themes from multiple angles with diverse phrasing; a low TTR indicates a more restricted vocabulary or frequent repetition of ideas using similar wording \cite{Savoy2020}.

    \subsection{The Impact of \textsc{TraceTarnish} on an Author's Stylistic Signature}
    \label{subsec:Impact_TraceTarnish}

        Now, we hypothesize why the \textsc{TraceTarnish} data yielded the observed results.

        On average, the anonymized \textsc{TraceTarnish} outputs were noticeably shorter than their non-anonymized input counterparts, primarily because of the adversarial obfuscation step. This step replaces key nouns, verbs, and modifiers with synonyms or alternative constructions, making it harder for keyword-based detectors to match the original phrasing while preserving the overall meaning.

        The adversarial translation step likely removes articles and replaces proper nouns and similar elements with functional words, reducing overall lexeme counts. The round-trip translation typically introduces lexical drift, grammatical shifts, or loss of nuance.

        Finally, the adversarial steganography step, presumably, affects the Type-Token Ratio (TTR) metric, as this aspect of the attack can virtually double the number of tokens registered for a given text. The zero-width Unicode characters are likely to be (A) interpreted as legitimate whitespace instances or (B) valid characters deserving of analysis, causing a word such as \say{privacy} to be parsed as (A) \say{pri} and \say{vacy} or (B) a string that visually resembles \say{privacy} but would not be detected as such\footnote{If robust sanitation protocols were applied to all input---whether LLM prompts or generic text fields---Unicode smuggling attacks that hide zero-width Unicode characters (similar to ASCII-smuggling attacks) would be largely neutralized. Proper filtering would detect and strip invisible code points before the text reaches downstream processors, preventing the hidden payload from being interpreted as executable or meaningful data.}.

        These alterations appear to disproportionately affect function words and content words alike. By substituting or removing articles, prepositions, and conjunctions, the adversarial steps reduce the frequency of common function-word types, which in turn lowers their contribution to the overall token count. Simultaneously, synonym replacements and lexical drift diversify the pool of content words, increasing the number of distinct content-word types while often shortening or fragmenting individual instances. The combined effect is a marked rise in the Type-Token Ratio: fewer repeated function words and a broader, more varied set of content words inflate the ratio, even though the total token count may decline. Consequently, the text becomes statistically more \say{lexically rich} but less faithful to the original distribution of function and content word categories.

        To illustrate our point, consider the following sentences. Original sentence: \say{If you are distressed by anything external, the pain is not due to the thing itself, but to your estimate of it; and this you have the power to revoke at any moment} (\href{https://archive.org/details/meditationsofmar00marc_0/page/112/mode/2up?q=%22If+thou+art+pained%22}{\textit{Marcus Aurelius}}). Adversarial version (illustrating the concept): \say{If you suffer for something external, the pain is not from him, but from his preciousness, and he has the power to withdraw at any time.}

\begin{figure}[H]
            \centering
            \subfloat{\includegraphics[width=0.50\linewidth]{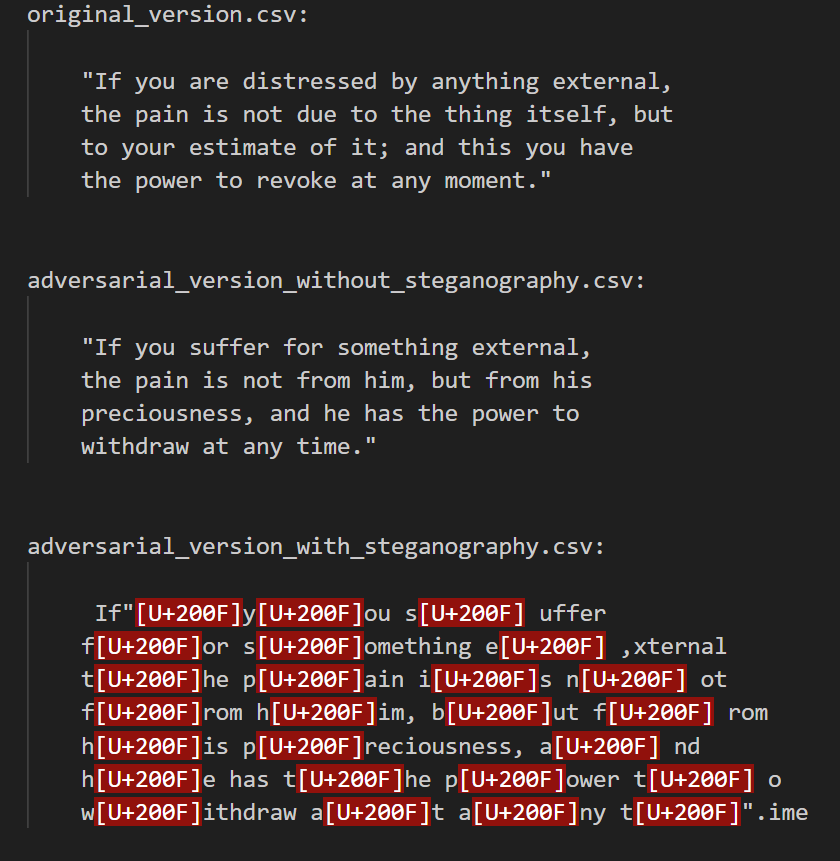}} \quad
            \subfloat{\includegraphics[width=0.80\linewidth]{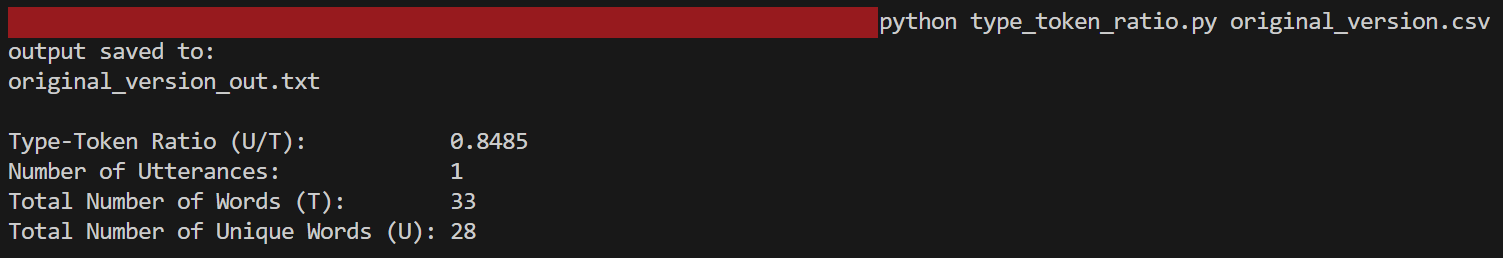}} \quad
            \subfloat{\includegraphics[width=0.80\linewidth]{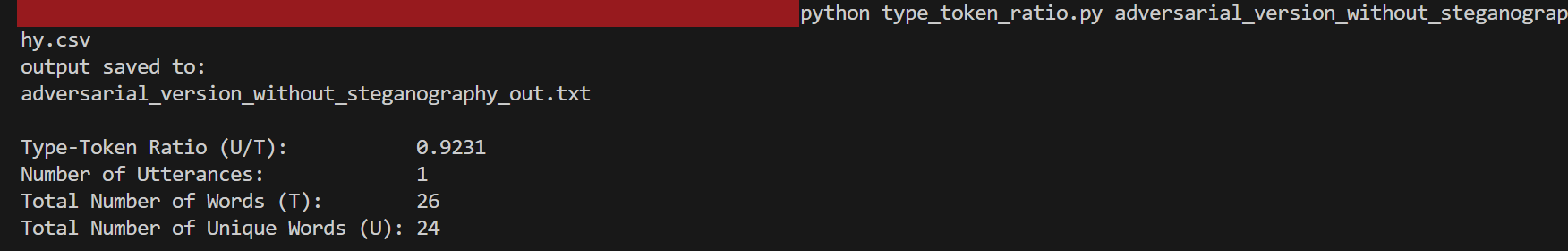}} \quad
            \subfloat{\includegraphics[width=0.80\linewidth]{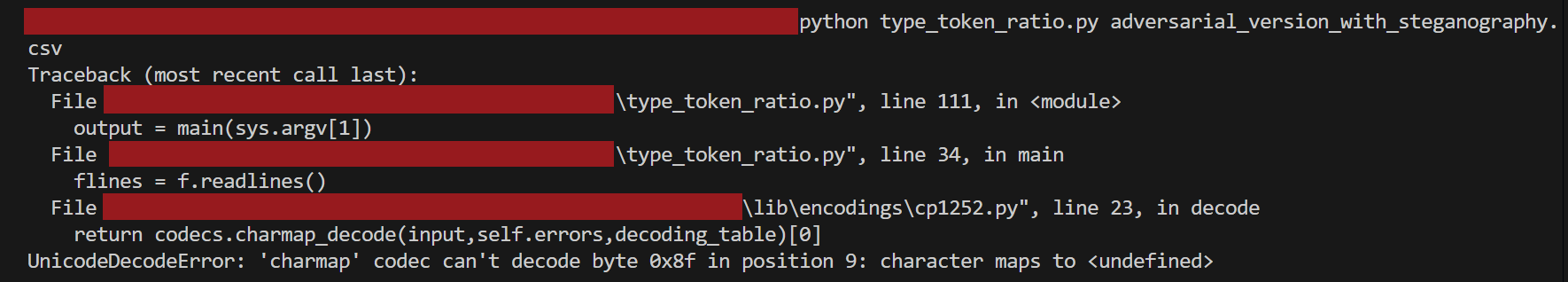}} \quad
            \subfloat{\includegraphics[width=0.80\linewidth]{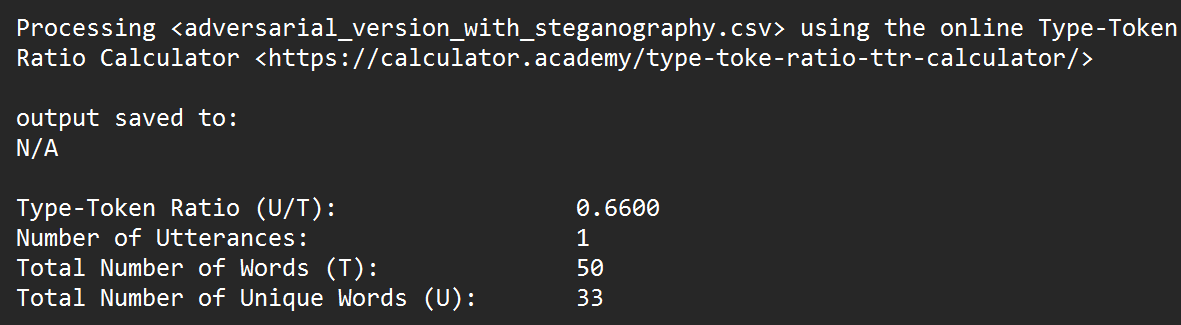}}
            \caption{The outputs of Steven C. Howell's ``type\_token\_ratio.py'' program \cite{Howell2016} using a processed and an unprocessed version of a Marcus Aurelius quote.}
            \label{fig:TTR_Calculation_Example}
        \end{figure}

        The text-analysis results for the original sentence are: TTR 0.848; unique words 28; total words 33. The results for the adversarial version \textit{without} steganographic tampering are: TTR 0.923; unique words 24; total words 26. From this, we can conclude that the edits increase the TTR, indicating a text that is statistically more lexically rich while using fewer overall tokens. The results for the adversarial version \textit{with} steganographic tampering are: TTR 0.660; unique words 33; total words 50. For clarification, the two adversarial versions look identical, at least visually, but the former aligns with our explanation, while the latter conforms to our experimental results. In the face of steganographic tampering, the TTR decreases and the word counts balloon; when all other adversarial steps are performed without steganography, the TTR increases and the word counts shrink. See (\textbf{Figure \ref{fig:TTR_Calculation_Example}}) for a visual aid of our TTR working example.

        Overall, \textsc{TraceTarnish} simplifies its textual input---producing a more simplistic rendition of the input text---while the text's semantic meaning is effaced through the incorporation of words that are valid in one language but invalid in another, and the number of tokens in the paraphrased text increases, roughly doubling the original count.

\section{Proposed Enhancements and Next Steps}
\label{sec:Proposed_Enhancements}

    With everything we have learned from our experimentation, it has become increasingly apparent that our attack pipeline would benefit from an \say{adversarial imitation} step, but one unlike our previously proposed implementation. Initially, we sequenced the components as Translation \( \rightarrow \) Imitation \( \rightarrow \) Obfuscation \( \rightarrow \) Steganography; the current iteration runs Translation \( \rightarrow \) Obfuscation \( \rightarrow \) Steganography. While this satisfies our ends, the quality of the text produced could always be improved. Much like Hume's guillotine (the is/ought problem)\footnote{Hume's guillotine, also called the is-ought problem, is the claim that one cannot logically derive prescriptive (\say{what ought be} or \say{how things should be}) statements solely from descriptive (\say{what is} or \say{observation-based}) premises. The gap between facts and values must be bridged by an additional normative premise.}, what \textit{ought} to be is a distant horizon, while what \textit{is}, is close and in immediate view. Our attack's outputs \textit{ought} to be sensible and comprehensible. Our attack \textit{is} effective in that it verifiably deteriorates an author's stylistic signature, with the trade-off being coherence. Although \textsc{TraceTarnish}'s outputs (more or less) embody the themes of their inputs, the state of the inputs after processing can have immense repercussions.

    A well-formed statement has a far better chance of producing a passable output, whereas colloquialisms, slang, and other aspects typical of conversational text can produce suboptimal results. This is expected; machines are deterministic, so analytically-driven processes such as machine translation and paraphrasing are also subject to that determinism. After all, the composition of genuinely authored text can be influenced by manifold influences and sundry stimuli; it is lithe, effervescent, and mercurial. This too appears in certain outputs, as non-conventionally structured text strewn with emoticons leads to unremarkable results.

    With all this in mind, we have concluded that some form of quality control is necessary, perhaps in the form of a large language model (LLM) \cite{Wang2025}. Given our particular ask, a lightweight LLM seems fitting. The idea would be to prompt an \textit{offline}, \textit{self-hosted} LLM to assume the role of Grammarly.

    The thought process behind our design is as follows. Our data should never leave our device, hence the \textit{self-hosting}. Our data, whatever it may be, should \textit{not} be used to train future models without our express consent, which is why the LLM will be \textit{offline}. If it were online, and we were to disclose our sensitive information in some chat, a log of that chat would be produced. That log could then be read and accessed by a third party; even if we deleted the original conversation, the \say{deleted} message may still be retained and probed for a variety of reasons \cite{Lightcap2025}.

    Our agent---or LLM instance with a clear objective---will also be tasked with retaining the steganographic payloads of the original text while proofreading and revising the input it receives.

    While performing its grammatical revisions, the LLM will be asked to randomly assume the identity (or profile) of a random person, randomizing gender, age, education level, and nationality. This subtask will characterize the code component as a whole, earning the \say{adversarial imitation} label, if only in spirit.

\begin{figure}[H]
        \centering
        \includegraphics[width=1\linewidth]{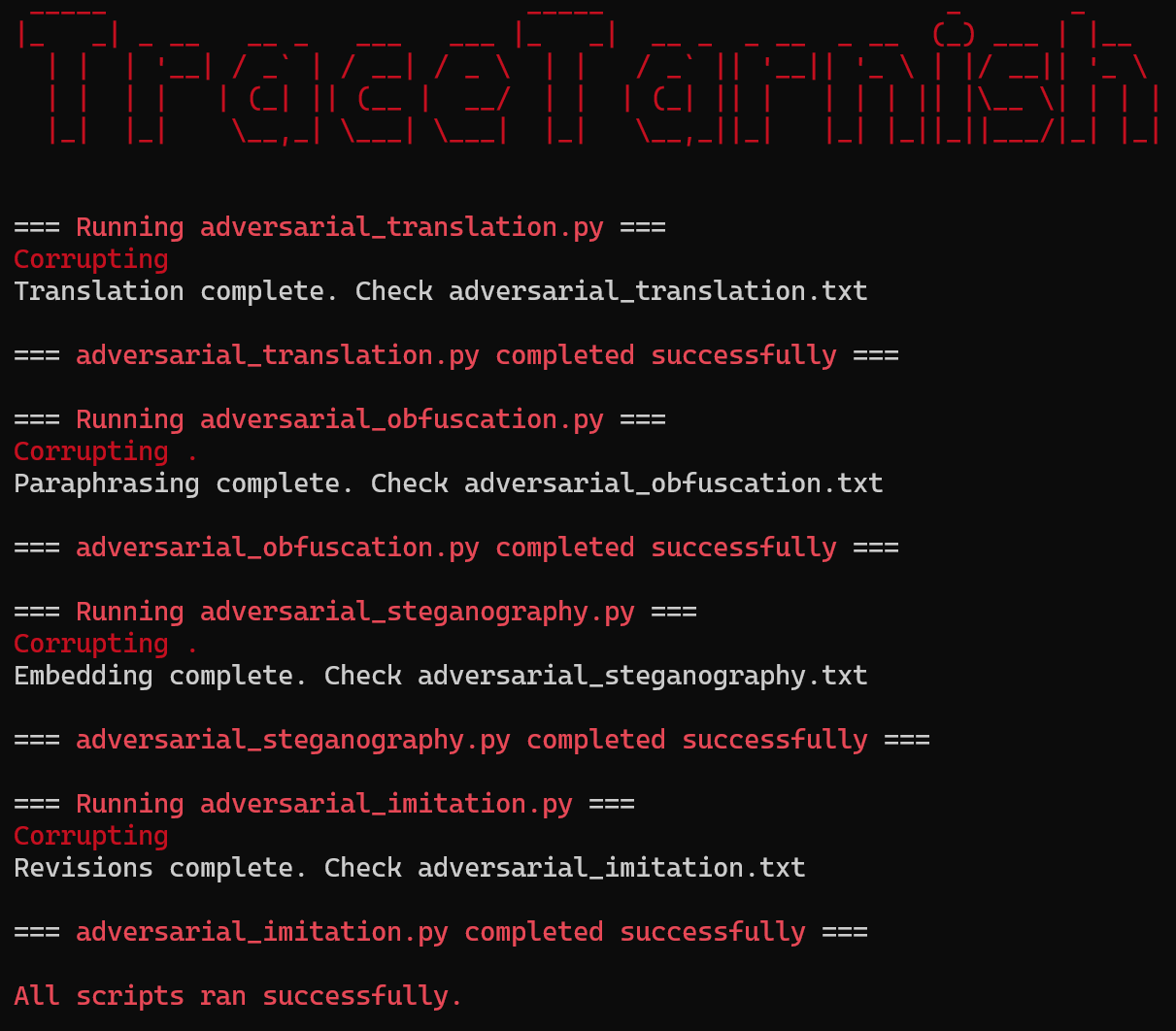}
        \caption{The terminal output for our updated \textsc{TraceTarnish} script.}
        \label{fig:TraceTarnish_Terminal_Output_v2}
    \end{figure}

\begin{figure}[H]
        \centering
        \includegraphics[width=1\linewidth]{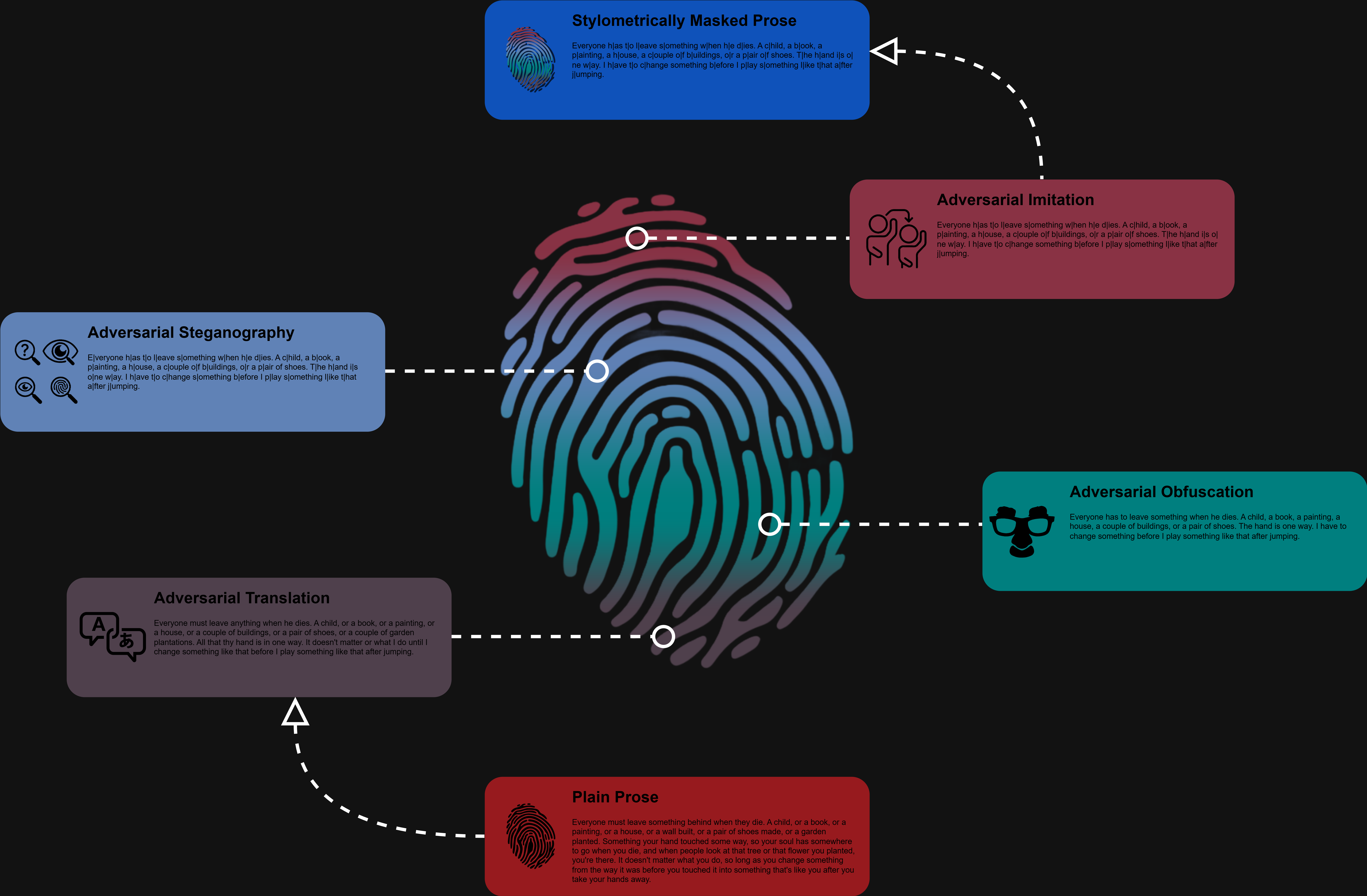}
        \caption{To bring something into existence---the very act of creation---imbues, imparts, and etches a fragment of your soul onto whatever you produce. This corporality---making something from nothing---is suffused with an undeniable essence---an essence that is uniquely yours. While sacred, that essence also erodes privacy in the most profound way: what is unique is detectable; what is detectable is leveragable. Our aim with this attack is to \textcolor{red_teaming}{\textit{"burn [this essence] to ashes, [and] then burn the ashes"}} (\textit{Ray Bradbury}). Pictured above is a sample workflow visualization of our updated \textsc{TraceTarnish} script, in which we process another of Bradbury's quotes.}
        \label{fig:TraceTarnish_Sample_Workflow_Visual_v2}
    \end{figure}

    Our attack sequence, with the above modifications, will become Translation \( \rightarrow \) Obfuscation \( \rightarrow \) Steganography \( \rightarrow \) Imitation. As before, we will refactor the code, add to it, and evaluate the changes iteratively. See (\textbf{Figures \ref{fig:TraceTarnish_Terminal_Output_v2}; \ref{fig:TraceTarnish_Sample_Workflow_Visual_v2}}) for our code's current implementation.

\section{Conclusion}
\label{sec:Conclusion}

    %=== Source: https://mirrors.dotsrc.org/ctan/macros/latex/contrib/shapepar/shapepar.pdf ===
    %\epigraph{\shapepar{\eyeshape}{\textcolor{red_teaming}{We need to be really bothered once in a while. How long is it since you were really bothered? About something important, about something real?}}}{\textit{Fahrenheit 451 \\ Ray Bradbury}}
    \epigraph{    
        \includegraphics[width=1\linewidth]{{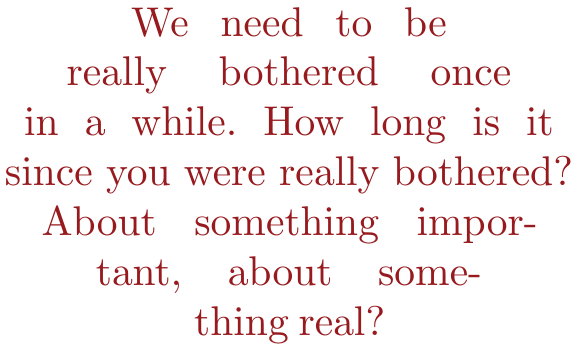}}
    }{\textit{Fahrenheit 451 \\ Ray Bradbury}}

    We would like to bring our current study to a close by asking two open-ended questions. First, what makes you, \textit{you}? This, naturally, invites a deeper, more disconcerting inquiry. How could the answer to the first question be used against \textit{you}?

    From the perspective of computational stylistics (stylometry), a robust profile can be synthesized using the counts of function words, tallies of function word types, and the richness of the author's vocabulary (or the proportion of unique words by the length of their write-up).

    Like it or not, \textit{you} can be reduced to a number. \textit{You} can be identified, tracked, and censored based on a numerical score or array of scores. This is \textit{you}, and what makes you, \textit{you} is exploitable. Does it have to be this way? Or is there a better way to be?

    While not perfect---and certainly not complete---\textsc{TraceTarnish} endeavors to eviscerate the \textit{you} that can be taken advantage of. But it is not without its shortcomings. As we've uncovered, calculating and exploring the correct subset of stylometric features may signal the utilization of \textsc{TraceTarnish}.

    Be that as it may, there is still work to be done, for the erosion of privacy is \textit{nigh} \cite{Discord2025}, surveillance is \textit{imminent} \cite{Belanger2025}, and censorship \textit{looms around the corner} \cite{Goldman2025}.

    Now, we leave you with a final question: what are \textit{you} going to do about it\footnote{\textcolor{red_teaming}{For you see, \say{\dots [we] are looking at\dots [those who are craven]. [We] saw the way things were going, a long time back. [We] said nothing. [We're] one of the [apathetic] who could have spoken up and out when no one would listen to the `[culpable],' but [we] did not speak and thus became [culpable ourselves].}} (\textit{Ray Bradbury})}?

\bibliographystyle{splncs04}
\bibliography{TraceTarnish.bib}

\end{document}